# Characterization of a naphthalene dioxygenase endowed with an exceptionally broad substrate specificity towards polycyclic aromatic hydrocarbons†

Yves Jouanneau [*,‡], Christine Meyer[‡], Jean Jakoncic[§], Vivian Stojanoff[§], Jacques Gaillard[||]

[‡]CEA, DSV, DRDC, Lab. Biochim. Biophys. Syst. Intégrés; CNRS, UMR 5092, F-38054 Grenoble, France ;

[§]Brookhaven National Laboratory, National Synchrotron Light Source, Upton, NY 11973, USA;

[||]CEA, DRFMC, SCIB, LRM, UMR UJF-CEA 3, F-38054 Grenoble , France.

Running title : A broad specificity ring-hydroxylating dioxygenase

† This work was supported by grants from the Centre National de la Recherche Scientifique, the Commisariat à l’Energie Atomique and the Université Joseph Fourier to UMR5092.

Corresponding author:

Yves Jouanneau

CEA-Grenoble, DRDC/BBSI,

F-38054 Grenoble Cedex 9, France.

Tel. : 33 (0)4.38 78.43.10; Fax : 33 (0)4.38 78.51.85

Email : yves.jouanneau@cea.fr

**Abbreviations**

GC-MS, gas chromatography coupled to mass spectrometry;

HEPES, N-(2-hydroxyethyl)piperazine-N'-(2-ethanesulfonic acid!;

HPLC, high performance liquid chromatography;

Ht, His-tagged;

IMAC; immobilized metal affinity chromatography;

IPTG, isopropyl-β-D-thiogalactopyranoside;

PAH; polycyclic aromatic hydrocarbon;

$\text{Red}_{B356}$, reductase component of the biphenyl dioxygenase from *C. testosteroni* B356;

RHD, ring hydroxylating dioxygenase

**Abstract**

In *Sphingomonas* CHY-1, a single ring-hydroxylating dioxygenase is responsible for the initial attack of a range of polycyclic aromatic hydrocarbons (PAHs) composed of up to five rings. The components of this enzyme were separately purified and characterized. The oxygenase component (ht-PhnI) was shown to contain one Rieske-type [2Fe-2S] cluster and one mononuclear Fe center per alpha subunit, based on EPR measurements and iron assay. Steady-state kinetic measurements revealed that the enzyme had a relatively low apparent Michaelis constant for naphthalene ($K_m$= 0.92 ± 0.15 μM), and an apparent specificity constant of 2.0 ± 0.3 $\mu M^{-1}$ $s^{-1}$. Naphthalene was converted to the corresponding 1,2-dihydrodiol with stoichiometric oxidation of NADH. On the other hand, the oxidation of eight other PAHs occurred at slower rates, and with coupling efficiencies that decreased with the enzyme reaction rate. Uncoupling was associated with hydrogen peroxide formation, which is potentially deleterious to cells and might inhibit PAH degradation. In single turnover reactions, ht-PhnI alone catalyzed PAH hydroxylation at a faster rate in the presence of organic solvent, suggesting that the transfer of substrate to the active site is a limiting factor. The four-ring PAHs chrysene and benz[a]anthracene were subjected to a double ring-dihydroxylation, giving rise to the formation of a significant proportion of *bis-cis*-dihydrodiols. In addition, the dihydroxylation of benz[a]anthracene yielded three dihydrodiols, the enzyme showing a preference for carbons in positions 1,2 and 10,11. This is the first characterization of a dioxygenase able to dihydroxylate PAHs made up of four and five rings.

Ring-hydroxylating dioxygenases (RHDs) are widely spread bacterial enzymes that play a critical role in the biological degradation of a large array of aromatic compounds, including polycyclic aromatic hydrocarbons (PAHs)(*1, 2*). RHDs catalyze the initial oxidation step of such compounds, which consists in the hydroxylation of two adjacent carbon atoms of the aromatic ring, thus generating a *cis*-dihydrodiol. This reaction converts hydrophobic, often toxic, molecules, into more hydrophilic products, allowing for their subsequent metabolism by other bacterial enzymes. Some RHDs were found to attack highly recalcitrant environmental pollutants, including dibenzo *p*-dioxin (*3, 4*), polychlorobiphenyls (*5*), and PAHs (*6-8*), thus promoting studies on this type of enzymes with the ultimate goal of improving bioremediation processes (*2, 9*). RHDs are multi-component enzymes, generally composed of a NADH-oxidoreductase, a ferredoxin and an oxygenase component that contains the active site. Sometimes, the reductase and the ferredoxin are fused in a single polypeptide. The oxygenase component is a multimeric protein, with either an $\alpha_n\beta_n$ (n=2 or 3) or $\alpha_3$ structure, that contains one [2Fe-2S] Rieske cluster and one non-heme iron atom per $\alpha$ subunit (*1*). During a catalytic cycle, two electrons from the reduced pyridine nucleotide are transferred, via the reductase, the ferredoxin and the Rieske center, to the Fe(II) ion at the active site. The reducing equivalents allow the activation of molecular oxygen, which is a prerequisite to dihydroxylation of the substrate (*10*).

So far, only a few RHDs have been purified and extensively characterized, including phthalate dioxygenase (*11, 12*), naphthalene dioxygenase (*13, 14*) and biphenyl dioxygenase (*15*). None of these enzymes is able to oxidize substrates with more than three fused rings, and data on the mechanism, kinetics and efficiency of the oxidation of high molecular weight PAHs by bacterial dioxygenases are relatively scarce (*16*). However, the four-ring PAHs chrysene and benz[a]anthracene, and the five-ring benzo[a]pyrene are of particular concern because they are well-documented carcinogens (*17*). Recently, a Sphingomonad endowed with the remarkable ability to grow on chrysene as sole carbon and energy source was isolated in our laboratory (*18*). In this strain, called *Sphingomonas* sp. CHY-1, a single dioxygenase was shown to be responsible for the oxidation of polycyclic hydrocarbons made of 2 to 4 rings (*6*). In the present study, the three components of the dioxygenase were purified and characterized, and the catalytic properties of the enzyme with respect to the oxidation of nine PAHs were examined. Due to the broad specificity of this enzyme, the kinetics and coupling efficiency of the dioxygenase-catalyzed reaction with 2 to 5-ring PAHs could be compared for the first time. Steady-state kinetic parameters were determined for representative 2-ring PAHs. In addition, the reactivity and regioselectivity of the enzyme towards benz[a]anthracene was further investigated by means of single turnover chemistry and EPR spectroscopy.

## MATERIALS AND METHODS

### *Bacterial strains and growth conditions*

Strains of *Escherichia coli* and *Pseudomonas putida* carrying the relevant expression plasmids, as well as general culture conditions, have been previously described (*6*). Large-scale cultures required for the purification of the enzyme components were grown on rich medium, either Luria-Bertani or Terrific broth (*19*), in a 12-L fermentor (Discovery 100, SGI-Inceltech/New Brunswick Scientific, Paris, France). Cultures destined to the overproduction of the oxygenase or the ferredoxin component were supplemented with 50 µM ferrous ammonium sulfate. The medium was inoculated with 400 ml of an overnight culture, then incubated at 37°C under constant aeration and agitation (500 rpm), until the bacterial density ($OD_{600}$) reached about 1.0. The temperature was then lowered to 25°C, IPTG was added to 0.2 mM final concentration, and the culture was further incubated for 20 h before being harvested by centrifugation. The bacterial pellet was washed with 50 mM Tris-HCl buffer (pH 7.5), and kept frozen until use.

### *Protein purification*

All purification procedures were carried out under argon, using buffers equilibrated for at least 24 h in a glove box maintained under anoxic conditions ($O_2$ <2 ppm, Jacomex , France). The temperature was kept at 0-4 °C except when otherwise indicated. Crude extracts were prepared by thawing the bacterial pellets in twice as much lysis buffer by volume, followed by lysozyme treatment (0.5 mg/ml) for 15 min at 30°C. The lysis buffer was either 50 mM Tris-HCl, pH 7.5 (oxygenase preparation), 50 mM Tris-HCl, pH 8.0, 0.5 M NaCl, 10% glycerol (reductase preparation) or 50 mM potassium phosphate, pH 7.5, 0.5 M NaCl, 10% glycerol, 2 mM β-mercaptoethanol (ferredoxin preparation). The suspension was then subjected to ultrasonication for a total time of 5 min at 80% of maximal intensity, using a Vibra Cell apparatus run in pulse mode at 5 s/pulse (Fisher Bioblock Scientific, Illkirch, France). The lysate was centrifuged at 12,000 g for 30 min, and the resulting cell extract was used as the starting material for protein purification.

***Purification of the oxygenase component PhnI***

A cell extract was prepared as described above from *P. putida* KT2442 carrying plasmid pSD9 (*6*). The extract obtained from approx. 50 g of cells was diluted two-fold with TGE buffer (25 mM Tris-HCl, pH 7.5, containing 5% glycerol, 5% ethanol, and 2 mM β-mercaptoethanol), and applied to a 40-ml column of DEAE-cellulose (DE52, Whatman) equilibrated with TGE buffer. After washing the column with four bed volumes of the same buffer, the oxygenase was eluted as a brown band with buffered 0.3 M NaCl. The eluate was immediately applied to a small column (7 ml) of immobilized metal affinity chromatography (IMAC) resin loaded with $Co^{2+}$ (TALON, BD Biosciences Ozyme, France). The column was washed successively with 8 bed volumes of TGE buffer containing 0.5 M NaCl, and 5 bed volumes of the same buffer supplemented with 20 mM imidazole. A brown protein fraction was then eluted with TGE buffer containing 0.15 M imidazole. This fraction was diluted 6-fold with TGE buffer and applied to a small column of DEAE-cellulose (4 ml). The purified protein was eluted in a small volume of TGE buffer containing 0.3 M NaCl, and frozen as pellets in liquid nitrogen. This preparation, designated ht-PhnI, was judged to be at least 95% pure by SDS-PAGE.

***Purification of the ferredoxin component PhnA3***

PhnA3 was overproduced in *E. coli* BL21AI (Invitrogen) carrying plasmid pEBA3 (*15*). The cell extract prepared from 154 g packed cells was loaded onto two columns of IMAC-TALON, 13 ml each, equilibrated in PG buffer (50 mM potassium phosphate, pH 7.5, 10% glycerol, 2 mM β-mercaptoethanol), containing 0.5 M NaCl. Each column was washed with 100 ml of equilibration buffer, followed by 50 ml of buffered 20 mM imidazole. A brown protein fraction was eluted with PG buffer containing 0.15 M imidazole. This fraction was immediately diluted 5-fold with PG buffer, and loaded onto a 10-ml DEAE cellulose column. After washing with two bed volumes of PG buffer, the brown ferredoxin fraction was eluted with buffered 0.3 M NaCl in a volume of 8.6 ml. This preparation was designated ht-PhnA3. Part of the purified His-tagged protein (4.7 µmoles) was cleaved by incubation with thrombin (10 U/ µmole) for 16 h at 20°C, in buffer containing 0.15 M NaCl and 2 mM $CaCl_2$, pH 8.2. The digested protein was passed through a 2-ml IMAC-TALON column, diluted 5-fold with PG buffer, then loaded onto a 2-ml column of DEAE-cellulose. The ferredoxin was eluted in a small volume PG buffer containing 0.3 M NaCl, and frozen as pellets in liquid nitrogen. This preparation was referred to as rc-PhnA3.

***Purification of the reductase component PhnA4***

PhnA4 was overproduced in *E. coli* BL21(DE3) carrying plasmid pEBA4 (*15*). The crude extract from 16 g of cells was applied to a 2-ml column of IMAC-TALON equilibrated in TG buffer (Tris-HCl, pH 8.0, 10% glycerol) containing 0.5 M NaCl. The column was washed with 10 bed volumes of equilibration buffer, and 3 bed volumes of TG buffer containing 0.5 M NaCl and 10 mM imidazole. A yellow protein fraction was then eluted with TG buffer containing 0.15 M imidazole. This fraction was dialyzed for 16 h against TG buffer, and further purified on a second column of IMAC-TALON (1 ml). The column was successively washed with TG buffer containing 0.5 M NaCl (10 ml), and the same buffer containing 10 mM (9 ml), and 20 mM (4 ml) imidazole. The reductase was eluted in a small volume of TG buffer containing 0.15 M imidazole, dialyzed as above, and concentrated to 0.8 ml by ultrafiltration using an Ultrafree centifugal device with 30-kDa cut-off (Millipore, Amilabo, France). The purified protein was stored as pellets in liquid nitrogen.

***Purification of ht-Red$_{B356}$***

The reductase component of the biphenyl dioxygenase from *C. testosteroni* B-356, designated as ht-Red$_{B356}$, was purified from *E. coli* SG12009(pREP4)(pEQ34*::bphG*) (*20*), as previously described (*21*).

***Enzyme assays***

Dioxygenase activity was assayed either by following NADH oxidation at 340 nm or by measuring the rate of $O_2$ consumption using a Clark-type $O_2$ electrode (Digital model 10; Bioblock Scientific, Illkirch, France). Polarographic measurements (standard assay) were performed at 30°C in reaction mixtures (1 ml) containing 0.13 µM ht-PhnI, 1.57 µM PhnA3, 0.40 µM ht-Red$_{B356}$ , and 0.5 mM NADH in 50 mM potassium phosphate buffer, pH 7.0. The PAH substrate was supplied at 0.1 mM from a concentrated solution in acetonitrile. The concentration of all three enzyme components was doubled for assays with 4- and 5-ring PAHs. The reaction was initiated by injecting, with a gas-tight syringe, a 100-fold concentrated mixture of the proteins kept under argon on ice in phosphate buffer containing 10% glycerol, 10 mM dithiothreitol and 0.05 mM ferrous ammonium sulfate. The enzyme activity was determined from the initial rate of $O_2$ consumption and expressed as µmol $O_2$ per min per mg ht-PhnI. Reaction rates were calculated from duplicate assays and corrected for the $O_2$ consumption measured in control assays carried out in the absence of PAH substrate.

The $O_2$ electrode was also used to determine the coupling between PAH oxidation and $O_2$ consumption as follows. After approx. 50 nmol $O_2$ had been consumed, 300 U of bovine liver catalase (Sigma) was added as a means to estimate the amount of $H_2O_2$ generated during the reaction. Then, 0.6 ml of reaction mixture was withdrawn and immediately mixed with an equal volume of ice-cold acetonitrile. The dihydrodiols present in these samples were directly quantified by HPLC as described below. When appropriate, the dihydrodiols were extracted with ethyl acetate, derivatized and analyzed by GC-MS (see below).

To determine the coupling efficiency between NADH and PAH oxidation, some reactions were carried out in Eppendorf tubes containing 0.1 mM of substrate and 0.2 mM of NADH in 0.6 ml of reaction mixture. In those assays, the concentrations of the enzyme components were 0.38 µM (ht-PhnI), 6 µM (PhnA3) and 2 µM (ht-$Red_{B356}$). After an incubation time of 2 to 10 min at 30°C, depending on the substrate, the reaction was stopped by addition of an equal volume of acetonitrile. Residual NADH, and the dihydrodiols formed during the enzymatic reaction, were separated and quantified by HPLC.

Steady-state kinetic parameters of the dioxygenase-catalyzed reaction were determined from sets of enzyme assays where the substrate concentration was varied over a 0.5-100 µM range. The component ratio was the same as in the standard assay, but the protein concentration in the assays was 1.67-fold higher. The initial NADH concentration was 0.2 mM. Reactions were carried out at 30°C in quartz cuvettes, and the absorption at 340 nm was recorded at 0.1-s intervals over 1 min with a HP8452 spectrophotometer (Agilent Technologies, Les Ulis, France). The enzyme activity was calculated from the initial linear portion of the time course, using an absorption coefficient of 6,220 $M^{-1}.cm^{-1}$ for NADH. When biphenyl was used as substrate, NADH oxidation was recorded at 360 nm ($\varepsilon_{360}$ = 4,320 $M^{-1}.cm^{-1}$), because biphenyl 2,3-dihydrodiol absorbed at 340 nm. All assays were performed in duplicate, and at least 12 concentrations were tested per substrate. Plots of the initial reaction rate versus substrate concentration were fitted to the Michaelis-Menten equation using the curve fit option of Kaleidagraph (Synergy Software). Only curve fits showing correlation coefficients better than 0.98 were considered.

***Single turnover reactions***

Ht-PhnI was diluted to 57 µM in 20 mM HEPES, pH 7.0 containing 10% glycerol and 5 µM methyl viologen under argon, then reduced with a stoichiometric amount of dithionite. The reduction was checked by monitoring the protein absorbance in the 300-600 nm range. A portion of the reduced protein (50 µL) was diluted in 0.55 ml of air-saturated HEPES buffer containing 0.1 mM of PAH substrate. In some experiments, the buffer also contained a proportion of acetonitrile, as indicated. After incubation at 30°C for up to 10 min, the reaction was stopped by mixing with an equal volume acetonitrile containing 0.8% acetic acid. The mixture was heated for 2 min at 90°C, centrifuged and subjected to HPLC analysis as described below. Part of the solution was also extracted with ethyl acetate, and analyzed by GC-MS.

***Identification and quantification of reaction products***

Determination of dihydrodiols and the residual NADH concentration at the end of the dioxygenase-catalyzed reactions was performed by HPLC using a Kontron system equipped with F430 UV detector. Samples (0.2 ml) were injected onto a 4×150-mm C8 reverse-phase column (Zorbax, Agilent Technologies, France) run at 0.8 ml/min. The column was eluted with water for 2 min, then with a linear gradient to 80% acetonitrile for 8 min, and finally with 80% acetonitrile for 5 min. Detection was carried out at 340 nm (for residual NADH), and one of the following wavelengths, which was varied as a function of the absorbance maxima of the PAH dihydrodiols: 220 nm (naphthalene), 303 nm (biphenyl), 260 nm (phenanthrene), 244 nm (anthracene), 263 nm (benz[a]anthracene), 278 nm (chrysene), 280 nm (benzo[a]pyrene). Quantification was performed on the basis of peak area using calibration curves obtained by injecting known amounts of each dihydrodiol. Residual NADH was determined from the peak eluting at 2.6 min. The three benz[a]anthracene dihydrodiols formed by ht-PhnI were not resolved under the HPLC conditions used, and were estimated as a sum of their individual contribution, given that their absorbance coefficients at 263 nm were close to 31,000 $M^{-1}.cm^{-1}$. Purified 1,2-dihydroxy-1,2-dihydrobenz[a]anthracene was used for HPLC calibration. The extent of oxidation of fluorene and fluoranthene by the dioxygenase was determined from HPLC measurements of the amount of residual substrate. Wavelengths used for their detection were 262 and 236 nm, respectively.

PAH oxidation products generated by PhnI were also analyzed by GC-MS. Ethyl acetate extracts of samples were dried on sodium sulfate, evaporated under $N_2$, and derivatized with bis(trimethysilyl)trifluoroacetamide :trimethylchlorosilane (99:1) from Supelco (Sigma-Aldrich), prior to GC-MS analysis using a HP6890/HP5973 apparatus (Agilent Technologies). Operating conditions were as previously described (*22*), and mass spectrum acquisitions were carried out either in the total ion current or the single ion monitoring mode.

***Determination of the iron content of proteins***

To extract iron from proteins, samples (150 µl) were treated with 2.5 N HCl for 30 min at 95°C, then diluted with 0.7 volume of water. Iron was reacted with bathophenanthroline disulfonate (Sigma-Aldrich), and the complex formed was assayed by absorbance measurements at 536 nm (*23*). Assays were performed in triplicates. A calibration curve was generated by assaying serial dilutions of a standard solution of ferric nitrate, containing 1g/L of iron (Merck).

***Protein analyses***

Routine protein determinations were performed using the Bradford assay (*24*), or the bicinchoninic acid reagent kit (Pierce) using bovine serum albumin as a standard. The protein concentration of purified preparations of ht-PhnI was determined by a modification of the biuret assay (*25*). The absorbance coefficient of ht-PhnI at 458 nm was calculated to be 12,500 $M^{-1}.cm^{-1}$, on the basis of the latter assay. The concentrations of ht-PhnA3 and rc-PhnA3 were estimated from absorbance measurement at 460 nm, using an absorbance coefficient of 5,000 $M^{-1}.cm^{-1}$. SDS-PAGE on mini-slab gels was performed as previously described (*26*). The molecular masses of purified ht-PhnI and rc-PhnA3 were determined by size-exclusion chromatography on HR 10/30 columns of Superdex SD200 and SD75, respectively (both from Amersham Biosciences). The columns were run at a flow rate of 0.2 ml/min and calibrated with the following protein markers: Ferritin (443 kDa), catalase (240 kDa), aldolase (150 kDa), bovine serum albumin (67 kDa), ovalbumin (43 kDa) and myoglobin (17 kDa), aprotinin (6.5 kDa), all from Sigma-Aldrich, and ferredoxin VI from *Rhodobacter capsulatus* (11.58 kDa; (*27*)).

***EPR spectroscopy***

Protein samples were adjusted to a concentration of 20-40µM (ht-PhnI) or 100-500 µM (PhnA3) in argon-saturated phosphate buffer, pH 7.0, containing 10% glycerol. The redox status of the protein sample was checked by recording the absorbance spectrum, and, when appropriate, the protein was fully oxidized by injecting stoichiometric amounts of air with a gas-tight syringe. Ht-PhnI-nitrosyl complexes were prepared in a glove-box under argon (Jacomex) by incubating 190 µl of protein sample with 10 µl of 20 mM diethylamine NO-NOate (Cayman Chemical, Interchim, France) for 15 min. Samples were then introduced into EPR tubes and frozen in liquid nitrogen. In some experiments, protein samples were preincubated for 10 min with a PAH (0.1 mM) or a dihydrodiol, prior to NO-NOate addition. For analysis of the Rieske clusters, protein samples were reduced with an excess of sodium dithionite (1 mM). Full reduction was checked by absorbance recording prior to transferring the samples in EPR tubes and freezing them in liquid nitrogen. Spectra were recorded at a temperature set between 4 and 20 K with an X-band EMX Bruker spectrometer equipped with an ESR900 liquid helium cryostat (Oxford Instruments). Spin quantification was performed by integrating the appropriate signal, and comparing the signal intensity to that of the [2Fe-2S] ferredoxin (FdVI) from *Rhodobacter capsulatus*, taken as a reference (0.1 mM; (*27*)). The iron content of this reference sample was checked by chemical assay. All EPR tubes were calibrated in diameter.

***Chemicals***

$NAD^+$, NADH, PAHs, and most other chemicals were purchased from Sigma-Aldrich (Saint-Quentin-Fallavier, France). The *cis*-dihydrodiols used in this study were prepared from cultures of *E. coli* recombinant strains overproducing the PhnI dioxygenase, and incubated with a PAH. The purification and characterization of the diol compounds will be described elsewhere. The dihydrodiol concentrations were calculated using the following absorption coefficients: $\varepsilon_{262}$ = 8,114 $M^{-1}.cm^{-1}$ for *cis*-1,2-dihydroxy 1,2-dihydronaphthalene (*28*); $\varepsilon_{252}$ = 38,300 $M^{-1}.cm^{-1}$ and $\varepsilon_{260}$ = 43,000 $M^{-1}.cm^{-1}$ for *cis*-3,4-dihydroxy 3,4-dihydrophenanthrene (*29*); $\varepsilon_{244}$ = 55,600 $M^{-1}.cm^{-1}$ and $\varepsilon_{287}$ = 17,000 $M^{-1}.cm^{-1}$ for *cis*-1,2-dihydroxy 1,2-dihydroanthracene (*29*); $\varepsilon_{278}$ = 57,650 $M^{-1}.cm^{-1}$ for *cis*-3,4-dihydroxy 3,4-dihydrochrysene (*30*); $\varepsilon_{280}$ = 66,500 $M^{-1}.cm^{-1}$ for *cis*-9,10-dihydroxy-9,10-dihydrobenzo[a]pyrene (*31*); $\varepsilon_{263}$ = 31,000 $M^{-1}.cm^{-1}$ for *cis*-1,2- dihydroxy-1,2-dihydrobenz[a]anthracene and $\varepsilon_{275}$ = 37,000 $M^{-1}.cm^{-1}$ for *cis*-10,11-dihydroxy-10,11-dihydrobenz[a]anthracene (*32*).

## RESULTS

### *Purification and properties of the oxygenase component PhnI*

The His-tagged oxygenase component of strain CHY-1 dioxygenase, hereafter referred to as ht-PhnI, was anaerobically purified from *P. putida* KT2442(pSD9) in three steps as described under Materials and Methods. The procedure yielded approx. 12 mg of purified protein per liter of culture. The oxygenase was also produced in *E. coli* BL21(DE3)(pSD9), but the purification resulted in a lower yield. In addition, strain BL21(DE3) always produced a variable amount of insoluble recombinant protein (inclusion bodies), which was not the case when using strain KT2442 as host (data not shown). The latter strain was therefore preferred for overproduction and subsequent purification of ht-PhnI. SDS-PAGE analysis revealed that the ht-PhnI preparation was at least 95% pure, and was composed of two subunits with apparent $M_r$ of 52.000 and 20.000 (Fig. 1), consistent with the molecular masses of the polypeptides deduced from relevant gene sequences (*6*). Purified ht-PhnI exhibited a molecular mass of approx. 200 kDa by gel filtration chromatography, indicating that it is an $\alpha_3\beta_3$ hexamer. The brown protein showed a UV-visible absorbance spectrum with maxima at 280, 458 nm and a shoulder near 570 nm (data not shown), which is typical of proteins containing Rieske-type [2Fe-2S] clusters. The absorbance coefficient at 458 nm was found to be 12,500 $M^{-1}.cm^{-1}$ on average, as calculated from the protein content of three independent preparations of ht-PhnI with a similar content of [2Fe-2S] cluster (see Table 1). In contrast to related oxygenases previously characterized, ht-PhnI did not show a well-defined absorption band near 325 nm, but instead a shoulder likely resulting from two poorly resolved absorption bands. The $A_{280}/A_{458}$ ratio was relatively high (26.9) compared to that of naphthalene dioxygenase (17.6, (*33*)), a feature which might be partly explained by the higher content of aromatic residues of PhnI (Trp and Tyr account for 2.70 and 4.46 % of the total number of residues in PhnI versus 2.18 and 3.73% in naphthalene dioxygenase). EPR analysis of the reduced protein gave a rhombic signal with apparent g values at 2.02, 1.92 and 1.71, which is characteristic of Rieske-type [2Fe-2S] clusters (data not shown).

The iron content of the oxygenase was found to vary between 1.73 and 2.55 Fe atoms per pair of αβ subunits depending on preparations (Table I). In order to estimate the proportion of iron in each metal center, the ht-PhnI preparations were subjected to two independent EPR measurements. Upon reaction with NO, ht-PhnI gave rise to the formation of an Fe(II)-nitrosyl complex which was detected as an heterogeneous S = 3/2 EPR signal near g=4 (see Fig. 3). Based on the integration of that signal, the occupation rate of Fe(II) at the active site of the enzyme was found to vary between 0.20 and 0.92 (Table I). On the other hand, the estimation of the ratio cluster/αβ, calculated from the integration of the S= 1/2 signal in fully reduced protein samples, yielded values ranging between 0.75 and 0.85. Remarkably, the iron content of the preparations calculated from the sum of the two EPR determinations was in fairly good agreement with the total iron found by chemical assay.

The specific activity of the dioxygenase increased as a function of its iron content, but no clear correlation was observed between activity and the occupation rate of the active site (Table 1). In addition, preincubation of ht-PhnI with ferrous ions under reducing conditions prior to enzyme assay resulted in a marginal increase of activity (data not shown).

***Purification of the ferredoxin and reductase components***

The ferredoxin component was anaerobically purified as a His-tagged recombinant protein by IMAC chromatography, and designated ht-PhnA3. The purification procedure described herein yielded about 20 mg of ferredoxin per liter of culture, when strain BL21AI was used as a host for expression. Lower yields were observed with strain BL21(DE3)(pEBA3). The preparation was >90% pure as judged from SDS-PAGE. Cleavage of the protein with thrombin, followed by two short purification steps, gave an essentially pure preparation containing a 12-kDa polypeptide (Fig.1). The molecular mass of this protein, referred to as rc-PhnA3, was 13.5 kDa by gel filtration, which was slightly higher than the theoretical mass of the polypeptide calculated from the *phnA3* gene sequence (11,225 Da, (*6*)), but indicated that the ferredoxin was monomeric. Both ht-PhnA3 and rc-PhnA3 exhibited absorbance spectra indicative of partial reduction upon isolation under anoxic conditions, but rapidly oxidized in air. In the oxidized state, the two preparations of ferredoxin had identical spectra, featuring absorbance maxima at 278, 325 and 460 nm. The iron content of the ht-PhnA3 and rc-PhnA3 preparations was estimated to be 1.5 and 1.7 mol/mol of ferredoxin, respectively. EPR analysis of the reduced ferredoxin gave a signal with g values at 2.02, 1.90 and 1.82, which integrated to 0.86 spin/molecule. Taken together, these data provide strong evidence that the ferredoxin component contains one Rieske-type [2Fe-2S] cluster.

The reductase component of the dioxygenase encoded by *phnA4* was overproduced as a 45 kDa polypeptide in *E. coli* BL21(DE3)(pEBA4). However, a large proportion of the recombinant protein accumulated in the cells as inclusion bodies, and this problem was not solved by changing the host strain, or by lowering the temperature during induction. Although a low level of the reductase was recovered from the soluble cell extract, the recombinant protein was purified as a His-tagged fusion (ht-PhnA4) by affinity chromatography (0.2 mg/L of culture). The isolated ht-PhnA4 protein was yellow in color, and showed an absorbance spectrum typical for a flavoprotein, with absorbance maxima at 375 and 450 nm. Attempts to overexpress the reductase in *P. putida*, either intact or as a His-tagged fusion, under conditions similar to those described for PhnI, were unsuccessful (data not shown). These observations suggested that PhnA4 was an unstable protein. For our studies on the catalytic activity of the dioxygenase, we replaced PhnA4 by the more stable component, $Red_{B356}$, of the biphenyl dioxygenase from *C. testosteroni (20)*. Enzyme assays performed under standard conditions showed that $Red_{B356}$ efficiently substituted for PhnA4, and titration experiments with increasing concentrations of the reductase indicated that the two isoforms had almost identical affinities for rc-PhnA3 (data not shown).

***Catalytic properties of the dioxygenase complex : dependence of activity on electron carrier concentrations***

Purified ht-PhnI catalyzed the oxidation of naphthalene to *cis*-1,2-dihydroxy-1,2-dihydronaphthalene, in a reaction that required the presence of the reductase ($Red_{B356}$) and ferredoxin (ht-PhnA3) components. When the reductase concentration was varied, at constant concentrations of the ferredoxin and the oxygenase, activity reached half-saturation for a reductase concentration of 0.05 µM (data not shown). When the ht-PhnA3 concentration was varied, half-saturation was obtained when ferredoxin was added to an approx. 14-fold molar excess over the oxygenase concentration (4.5 µM). rc-PhnA3 was found to be equally active, indicating that the His-tag did not alter the enzyme function. At a ferredoxin concentration close to saturation (20 µM; 60-fold molar excess), the specific activity of the enzyme complex was calculated to be 1.25 ± 0.04 U/mg ht-PhnI. For most of the assays performed in this study, the reductase and ferredoxin concentrations were set at concentrations 2.4-fold and 12-fold higher than that of PhnI, respectively. A suboptimal level of ferredoxin was chosen to limit non-specific NADH oxidation by the protein mixture in the absence of PAH substrate.

***Specific activity and coupling efficiency***

To examine the ability of the dioxygenase to oxidize PAHs, the enzyme activity was first determined by measuring the initial rate of oxygen consumption in the presence of an excess of substrate. The dihydrodiol products formed in the reaction mixture were quantified by HPLC as described under Materials and Methods. When fluorene and fluoranthene were tested, substrate oxidation was rather estimated by measuring the amount of residual PAH at the end of the enzymatic reaction, because the oxidation products of these PAHs have not yet been fully characterized (see below and Table 3). In a second and independent set of experiments, the coupling efficiency of the PAH oxidation reactions catalyzed by the dioxygenase was determined by measuring the rates of NADH oxidation and dihydrodiol formation during catalysis. Table 2 compares the results obtained for nine PAHs in terms of specific activity, and reaction coupling between oxygen consumption, NADH oxidation and dihydrodiol formation.

Naphthalene appeared to be the only substrate yielding a stoichiometry close to 1, indicating a tight coupling between substrate and cofactor oxidation. It was also the best substrate as it gave the highest rates of $O_2$ consumption or NADH oxidation. Other substrates were utilized at rates that decreased with the number of fused rings, in reactions that gave rise to significant uncoupling between NADH oxidation and dihydrodiol formation. Chrysene appeared to be the worst substrate in terms of both oxidation rate and coupling efficiency. Although discrepancies were observed with some substrates when comparing enzyme activities assayed by $O_2$ consumption and NADH oxidation, the coupling efficiencies calculated as either dihydrodiol/$O_2$ or dihydrodiol/NADH ratios were similar within experimental error, except for phenanthrene and anthracene. It is unclear why different ratios were obtained in the two latter cases.

During steady-state catalysis, hydrogen peroxide was produced, with a $H_2O_2/O_2$ ratio that increased with the uncoupling of the reaction (Table 2). Depending on the substrate, the fraction of oxygen utilized for dihydrodiol formation varied between 8% (chrysene) and 100% (naphthalene), the balance of $O_2$ consumed being mainly allocated to $H_2O_2$ formation. However, some peroxide was produced even in the tightly coupled naphthalene hydroxylation reaction, the amount of which corresponded to the background $O_2$ consumption observed in the absence of PAH. Since the enzyme was saturated with naphthalene, the involvement of ht-PhnI in $H_2O_2$ formation was unlikely, suggesting that the electron carriers were responsible for this side reaction. In a control experiment, we observed that electron carriers alone gave rise to an $O_2$ consumption of 3.9 nmol.min$^{-1}$ compared to 5.1 nmol.min$^{-1}$ for the complete enzyme system, and generated 0.42 $H_2O_2$ per $O_2$ consumed (versus 0.57 for the complete system). Hence, a large proportion of the peroxide produced during *in vitro* catalysis of PAH hydroxylation was contributed by the electron carriers alone, most likely through air-oxidation of the reduced PhnA3 ferredoxin component.

***Steady-state kinetics***

Using naphthalene and biphenyl as substrates, the steady state rate of the PhnI-catalyzed reaction was determined in the 0.5-100 µM concentration range. The reaction was monitored spectrophotometrically, by measuring the kinetics of NADH oxidation. This assay method was preferred to the polarographic method, since at low substrate concentrations, the response time of the oxygen electrode was too long to account for the rapid consumption of the substrate. The dioxygenase exhibited a Michaelis-type behavior with respect to substrate concentration, and results indicated that the enzyme had an apparent $K_m$ as low as 0.92 ± 0.15 µM for naphthalene. The apparent turnover number for this substrate was 1.82 ± 0.03 s$^{-1}$. The enzyme showed a similarly low $K_m$ for biphenyl (0.42 ± 0.20 µM), the latter value being only an estimate as enzyme kinetics were extremely short (<4 s) and difficult to calculate accurately at substrate concentrations below 1.0 µM. The turnover number, expressed in terms of rate of dihydrodiol formed, was smaller (1.01 ± 0.04 s$^{-1}$), taking into account a dihydrodiol/NADH ratio of 0.67 in the calculation (Table 2). The apparent specificity constant was calculated to be 2.0 ± 0.3 µM$^{-1}$ s$^{-1}$ for naphthalene, and 2.4 ± 1.0 µM$^{-1}$ s$^{-1}$ for biphenyl.

***Dihydroxylations and monohydroxylations catalyzed by PhnI***

GC-MS analysis of the PAH oxidation products revealed that a single dihydrodiol was generated by the dioxygenase in most cases, except when fluorene, fluoranthene, chrysene and benz[a]anthracene were used as substrates (Table 3 and Fig. 2). Biphenyl, naphthalene, phenanthrene were hydroxylated at positions 2,3-, 1,2- and 3,4-, respectively. as previously determined (*6*), whereas anthracene was most likely converted to the 1,2-dihydrodiol, as found for the dioxygenase present in *S. yanoikuyae* B1 (*29*). Fluorene oxidation gave rise to the formation of five detectable products, four of which had mass spectra corresponding to monohydroxylated derivates (Table 3). While 9-fluorenol resulted from a monohydroxylation, the other products might have arisen from either a single hydroxylation or spontaneous dehydration of unstable dihydrodiols primarily produced by the enzyme, as proposed in a previous study on fluorene oxidation by naphthalene dioxygenase (*34*). Fluoranthene oxidation yielded only one detectable product with a mass spectrum characteristic of a monohydroxylated molecule (the prominent fragment at m/z=290 in Table 3 corresponds to the mass of the trimethylsilyl derivate of hydroxyfluoranthene), and a UV absorbance spectrum identical to that of 8-hydroxyfluoranthene (*35*). This result suggested that the dioxygenase catalyzed a monohydroxylation of fluoranthene on the C8 position. With chrysene, the major product detected was the *cis*-3,4-dihydrodiol, as determined by comparison of GC-MS and UV absorption data with those of the previously characterized diol (*30*). A more polar compound was also detected by HPLC, which accounted for less than 10% of the total products based on peak area. This compound, which gave a trimethysilyl derivate with a mass of 584 (Table 3), had the same chromatographic properties as the 3,4,9,10-*bis*-*cis*-chrysene dihydrodiol. We have independently identified this product based on proton and $^{13}C$ NMR (Jouanneau, Meyer, and Duraffourg, unpublished results). Finally, the dioxygenase-catalyzed oxidation of benzo[a]pyrene yielded a single product with a mass spectrum characteristic of a dihydrodiol derivate (Table 3). The UV spectrum of this product was identical to that of the *cis* -9,10-benzo[a]pyrene dihydrodiol (*31*).

***Dihydroxylation of benz[a]anthracene***

Benz[a]anthracene was converted by CHY-1 dioxygenase to three *cis*-dihydrodiol isomers and one *bis*-*cis*-dihydrodiol (Table 3 and Fig. 2). The three dihydrodiols have been independently purified and identified as the 1,2-, 8,9- and 10,11-isomers, based on a good match of GC-MS and UV absorbance data with previously published data (*32*). Quantitative analysis of the diols by GC-MS in several experiments showed that the 1,2-isomer was most abundant (68 ± 7%), with the 8,9- and the 10,11-isomers representing 9 ± 3% and 23 ± 4 % of the diols formed, respectively (average of 6 determinations). The proportion of *bis*-*cis*-dihydrodiol increased during the course of the enzymatic reaction, suggesting that at least one of the dihydrodiol reacted a second time with the enzyme to form the *bis*-*cis*-dihydrodiol. To test this hypothesis, the 1,2- and 10,11-isomers were independently provided as substrates to the dioxygenase. Interestingly, the two dihydrodiols triggered a fast and uncoupled oxidation of NADH, with small amounts of *bis*-*cis*-dihydrodiol produced (Table 2). Nevertheless, since the two isomers yielded the same product as judged from HPLC and GC-MS analysis, it is inferred that the *bis*-*cis*-dihydrodiol bore hydroxyls on carbons in positions 1, 2, 10 and 11 of the benz[a]anthracene molecule. The enzymatic reaction generated hydrogen peroxide at a rate much higher than that attributed to the electron carriers, indicating that, in this case, the formation of $H_2O_2$ was mainly due to futile cycling of the oxygenase.

***Reactivity of ht-PhnI toward benz[a]anthracene as investigated by single turnover experiments***

To further investigate the reactivity of the dioxygenase toward benz[a]anthracene, single turnover reactions were carried out under conditions similar to those previously described for naphthalene dioxygenase (*10*). In these experiments, the oxygenase component alone was allowed to react with the substrate in air-saturated buffer, and a rapid formation of dihydrodiol was expected at the enzyme active site. In a control experiment with naphthalene as substrate, the formation of dihydrodiol was observed on a time scale lower than 1 min. Surprisingly, the conversion of benz[a]anthracene was much slower and reached completion only after approx. 20 min. In addition, *bis-cis*-dihydrodiol was detected and its concentration increased linearly during the course of the reaction. These results suggested that the rate of the reaction was limited by the solubility of the substrate, which in turn reduced the accessibility of the substrate to the enzyme active site. On the other hand, the dihydrodiols produced which are soluble in water, might compete with the PAH for enzyme active sites, thus explaining the formation of *bis-cis*-dihydrodiol. This interpretation was tested in experiments where the solubility of the PAH was increased by adding an organic solvent to the reaction (Table 4). By carrying out the reaction in 20% acetonitrile, the solubility of benz[a]anthracene was increased 100-fold, and the reaction was completed in less than 1 min. In 30% acetonitrile, the reaction was also fast, but the product yield was lower, probably because of enzyme inactivation. GC-MS analysis of the diols formed showed that the 10,11-isomer was most abundant, and no *bis-cis*-dihydrodiol was detectable in reactions carried out in the presence of solvent. These results contrasted with those obtained in steady-state experiments, since in the latter case, the 1,2-isomer was the predominant product. However, the two sets of data could be reconciled by assuming that, under steady state conditions, the 10,11-isomer is converted to *bis-cis*-dihydrodiol faster than the 1,2-isomer (see below). The results of single turnover experiments demonstrate that the oxygenase preferentially hydroxylates benz[a]anthracene on carbons in positions 10,11. It is also shown that the reactivity of the enzyme towards water-insoluble substrates, which is limited by substrate transfer to the active site, could be enhanced by carrying out the reaction in aqueous medium containing up to 20% organic solvent.

***Interaction of ht-PhnI with benz[a]anthracene and dihydrodiols as probed by EPR spectroscopy***

The high level of uncoupling observed when benz[a]anthracene dihydrodiols were incubated with the dioxygenase indicated that these compounds did not interact correctly with the enzyme active site. As a means to probe this interaction, we carried out EPR analysis of complexes between the active site Fe(II) and NO, in the presence or absence of substrate (Fig. 3). The spectrum of the substrate-free enzyme showed a complex signal centered at g=4.0, which might reflect the existence of more than one NO binding site. Alternatively, the signals might arise from different conformations due to different orientations of the Fe-NO bond. Interestingly, two pairs of resonance lines at 3.68/4.40 and 3.98/4.07, which were prominent in the spectrum of the free enzyme, underwent dramatic changes upon substrate binding (Fig. 3). These lines almost disappeared in the spectrum of the benz[a]anthracene-bound enzyme, and were undetectable in the case of the naphthalene-bound enzyme complex. The shape of the signals obtained for the dihydrodiol-bound enzyme complexes showed patterns intermediate between the substrate-free and the benz[a]anthracene-bound enzyme, the spectrum of the 10,11-isomer-bound enzyme being closer to the latter. These observations could be taken as indirect evidence that the dihydrodiols were not correctly oriented in the substrate-binding pocket to allow for a productive catalytic conversion into *bis-cis*-dihydrodiol. Based on these EPR data, the 10,11-isomer would bind the active site in a more favorable position than the 1,2-isomer. Alternatively, our data might indicate that the enzyme was not saturated by the dihydrodiols, although concentrations in molar excess over the enzyme catalytic sites were used. Accordingly, in experiments where the concentration of the dihydrodiols was doubled (0.2 mM), we observed that the relative intensity of the resonance lines at 3.68/4.40 and 3.98/4.07, attributed to the free enzyme (see above), was significantly reduced compared to that found in spectra b and c in Fig. 3. The spectra of the enzyme-dihydrodiol complexes were then only slightly different from those obtained with naphthalene or benz[a]anthracene as ligands (data not shown). We therefore conclude that the differences seen in spectra b and c, compared to spectrum d, are likely due, in great part, to the contribution of substrate-free enzyme, thus reflecting a relatively low affinity of the enzyme for the two benz[a]anthracene dihydrodiols.

**DISCUSSION**

The ring-hydroxylating dioxygenase described in this study exhibits one of the broadest substrate specificities toward PAHs ever reported. It is a rare example of an enzyme able to attack aromatic substrates composed of 2 to 5 rings, which gave us an opportunity to compare the kinetics of dioxygenation for a wide range of PAHs. Consistent with previous *in vivo* observations (*6*), the specific activity of strain CHY-1 dioxygenase was highest with naphthalene, and declined as a function of substrate size. Ironically, while strain CHY-1 was isolated for its ability to grow on chrysene as sole carbon source (*18*), chrysene appeared to be the worst substrate in terms of both oxidation rate and coupling efficiency (Table 2). This finding suggested that a dioxygenase other than PhnI might be responsible for the initial attack of chrysene in strain CHY-1. However, such a hypothesis can be ruled out on the basis of our previous work showing that a mutant strain lacking PhnI failed to grow on and oxidize chrysene, as well as any other PAH used as substrate by the parental strain CHY-1 (*6*). Hence, PhnI is essential for growth on chrysene, and its low activity towards this substrate is probably one of the main reason why strain CHY-1 shows slow growth and poor cell yields when provided with chrysene as sole carbon source (*18*). In comparison, benz[a]anthracene, another 4-ring PAH which was oxidized at a much higher rate, failed to support growth of strain CHY-1 (*18*). Possible reasons which might explain this paradox are discussed below. The present study also unveiled interesting new features of strain CHY-1 dioxygenase, including its ability to utilize fluoranthene and benzo[a]pyrene. Hence, this enzyme has the remarkable potential to initiate the degradation of at least half of the 16 EPA priority PAHs, including the carcinogenic benz[a]anthracene and benzo[a]pyrene. A dioxygenase activity with a similarly broad substrate specificity has only been found in *S. yanoikuyae* B1, but the corresponding enzyme has not yet been described (*36*).

The dioxygenase from strain CHY-1 is a three-component enzyme that shares many of the biochemical properties of counterparts found in other bacteria degrading aromatic hydrocarbons. Based on the properties of the associated electron carriers, it would belong to class IIB of the dioxygenase classification proposed by Batie et al. (*37*), together with benzene and biphenyl dioxygenases (*1*). However, amino acid sequence analysis of the PhnI α and β subunits rather indicated that the enzyme was more closely related to naphthalene and phenanthrene dioxygenases (*6*), consistent with the substrate specificity of the dioxygenase determined herein. Purified ht-PhnI contained Rieske [2Fe-2S] and mononuclear Fe (II) centers as expected, but quantitative analysis based on EPR spectroscopy revealed that the two types of metal binding sites were not fully occupied. We cannot rule out metal loss during purification, despite employing conditions likely to minimize such losses. However, it is also possible that the biosynthesis of the oxygenase in *Pseudomonas* recombinant cells yielded a protein that did not have its full content of metal centers. Examples of purified oxygenases having a full complement of iron have occasionally been reported (*10, 38, 39*), but enzyme preparations partially lacking iron are more frequently obtained (*4, 40-42*). Hence, our EPR-based method to determine the occupancy of both iron binding sites might be of general interest for the characterization of such types of enzymes.

The *in vitro* activity of the dioxygenase was highly dependent on the component ratio ferredoxin over oxygenase, half saturation occurring for a 14-fold molar excess of the ferredoxin. Likewise, a high molar excess of ferredoxin was required to reach maximal activity in the case of naphthalene (*10*), and biphenyl dioxygenase (*39*). As a consequence, comparison of the apparent specific activities or $k_{cat}$ of enzymes from different sources should be regarded with caution. At a ferredoxin/oxygenase ratio of 14, the CHY-1 dioxygenase showed an apparent $k_{cat}$ of $1.82 \pm 0.03$ $s^{-1}$ and a specificity constant of $2.0 \pm 0.3$ $\mu M^{-1}$ $s^{-1}$ with naphthalene as substrate. A velocity constant 2.5-fold as high (4.48 $s^{-1}$) was found at a molar ratio of 60. In comparison, the biphenyl dioxygenase from *C. testosteroni* exhibited a $k_{cat}$ of $7.0 \pm 0.2$ $s^{-1}$ and a specificity constant of $1.2 \pm 0.1$ $\mu M^{-1}$ $s^{-1}$, at a ratio of 23 (*39*). An apparent $k_{cat}$ of 2.4 $s^{-1}$ and a specificity constant of 7.0 $\mu M^{-1}$ $min^{-1}$ (equivalent to 0.11 $\mu M^{-1}$ $s^{-1}$), was reported for 2-nitrotoluene dioxygenase from *Comamonas* JS765, at a ratio of 3.7 (*42*).

The present study revealed that the coupling between substrate oxidation and $O_2$ (or NADH) utilization varied widely depending on PAHs. While the conversion of naphthalene to dihydrodiol by the dioxygenase involved a stoichiometric amount of $O_2$, the oxidation of all other PAHs, except fluorene, gave rise to significant uncoupling. This uncoupling was associated with the release of $H_2O_2$, although a large proportion of the peroxide could be attributed to auto-oxidation of the electron carriers alone. Assuming that most of the peroxide detected in our *in vitro* assays is an artifact due to the great molar excess of electron carriers used in these assays, it is unclear whether peroxide would be produced in significant amounts *in vivo* as a consequence of the dioxygenase-catalyzed oxidation of PAHs, given that electron carriers are certainly in limiting amounts in natural host cells. Nevertheless, because of this uncoupling, part of the energy recovered from the catabolism of PAHs as NAD(P)H, is probably lost as unproductive transfer of reducing equivalents to $O_2$. This energy burden might affect growth yield, and could explain, at least in part, the higher resistance of 4- and 5-ring PAHs to bacterial biodegradation, inasmuch as the coupling efficiency of the dioxygenase reaction was lowest with those PAHs. Other dioxygenase systems were found to give rise to partially uncoupled reactions when challenged with poor substrates, and this was associated to a release of $H_2O_2$. Uncoupling occurred when naphthalene dioxygenase was incubated with benzene, and the peroxide formed was found to irreversibly inactivate the enzyme, probably because of the damage done by the product of the Fenton reaction between peroxide and the active site Fe(II) (*33*). Biphenyl dioxygenase also catalyzed uncoupled reactions and $H_2O_2$ production when provided with certain dichlorobiphenyls, which might result in the inhibition of the dioxygenation of other chlorobiphenyls. This effect, combined with the deleterious action of peroxide on cells, was predicted to inhibit the microbial catabolism of polychlorobiphenyls (*39*). Interestingly, some PAH-degrading bacteria were shown to specifically induce a catalase-peroxidase when grown on PAHs, thereby providing a means to cope with the dioxygenase-mediated formation of peroxide (*43*).

The oxidation of benz[a]anthracene by the dioxygenase from strain CHY-1 is of particular interest because this PAH was converted into three dihydrodiols, two of which were subjected to a second dihydroxylation in a highly uncoupled reaction. The amount of *bis-cis* dihydrodiol formed in this secondary reaction within the time of an assay (around 5 min) was estimated to be between 20 and 40 % of the total amount of diols primarily produced by the enzyme. These observations have several implications as for the reactivity and the coupling efficiency of the enzyme with respect to benz[a]anthracene. The rapid accumulation of *bis*-dihydrodiol indicated that the 1,2- and 10,11-dihydrodiols competed with benz[a]anthracene for the enzyme active site, which they reached faster than the PAH because of their much higher water solubility. This interpretation is supported by single turnover experiments showing that organic solvent accelerated benz[a]anthracene oxidation and suppressed *bis*-dihydrodiol formation. Because the dihydrodiols are poor substrates for the dioxygenase, as confirmed by EPR probing of nitrosyl-enzyme complexes, the competition they exert on benz[a]anthracene oxidation was expected to alter the coupling efficiency. A ratio dihydrodiol/$O_2$ of 0.31 was calculated without taking into account the formation of *bis*-dihydrodiol (Table 2). A calculation of the $O_2$ consumed for *bis*-dihydrodiol formation, assuming that 2 $O_2$ molecules were required per each molecule formed, allowed to bracket the ratio (*bis*-dihydrodiol + dihydrodiols)/$O_2$ between 0.42 and 0.55. This is definitely higher than the dihydrodiol/$O_2$ ratios found for chrysene and benzo[a]pyrene. Hence, based on coupling efficiency and oxidation rates, benz[a]anthracene appeared to be a better substrate than chrysene, and yet, it could not support growth of strain CHY-1. In *S. yanoikuyae* B1, a strain which cannot grow on benz[a]anthracene either, previous studies showed that the oxidation of this PAH led to the accumulation of three metabolites identified as 1-hydroxyanthranoic acid, 2-hydroxy 3-phenanthroic acid and 3-hydroxy 2-phenanthroic acid (*44*). These metabolites were predicted to arise from five similar degradation steps of benz[a]anthracene, involving an initial dihydroxylation on positions 1,2-, 8,9- and 10,11-, respectively. The data suggested that a subsequent step in the catabolic pathway of this PAH might be too slow to allow efficient processing of the metabolites, thereby preventing bacterial growth. Alternatively, the possibility was considered that the peroxide formed in the dioxygenase-catalyzed oxidation of benz[a]anthracene, or in the secondary oxidation of dihydrodiols, could inhibit growth. The secondary reaction generating *bis*-dihydrodiol is highly uncoupled, but it is unknown whether it would occur *in vivo* in strain CHY-1. *S. yanoikuyae* B1 did not produced any detectable *bis*-dihydrodiol when degrading benz[a]anthracene (*44*), indicating that dihydrodiols were rapidly metabolized. Accordingly, we have recently characterized a dihydrodiol dehydrogenase which can efficiently convert the three dihydrodiol isomers of benz[a]anthracene to corresponding catechols (*22*). Hence, in PAH-degrading Sphingomonads, the coupling between the first and the second enzymatic step of the catabolic pathway likely prevents the dioxygenase from catalyzing unproductive and potentially deleterious reactions.

In this work, we have purified and characterized a ring-hydroxylating dioxygenase with an exceptionally broad substrate specificity, which provides a good model for further structure-function studies on this class of enzymes. The oxygenase component has been recently crystallized and subjected to X-ray diffraction analysis. The structure of the protein has been solved to 1.85 Å resolution and will be described elsewhere (J. Jakoncic, Y. Jouanneau, C. Meyer, V. Stojanoff, unpublished data).

## ACKNOWLEDGMENTS

We thank John Willison for helpful discussions and critical reading of the manuscript.

## REFERENCES

(1) Butler, C. S., and Mason, J. R. (1997) Structure-function analysis of the bacterial aromatic ring- hydroxylating dioxygenases. *Adv. Microb. Physiol. 38*, 47-84.

(2) Gibson, D. T., and Parales, R. E. (2000) Aromatic hydrocarbon dioxygenases in environmental biotechnology. *Curr. Opin. Biotechnol. 11*, 236-243.

(3) Armengaud, J., Happe, B., and Timmis, K. N. (1998) Genetic analysis of dioxin dioxygenase of *Sphingomonas* sp. Strain RW1: catabolic genes dispersed on the genome. *J. Bacteriol. 180*, 3954-3966.

(4) Bunz, P. V., and Cook, A. M. (1993) Dibenzofuran 4,4a-dioxygenase from *Sphingomonas* sp. strain RW1: angular dioxygenation by a three-component enzyme system. *J. Bacteriol. 175*, 6467-6475.

(5) Furukawa, K., Suenaga, H., and Goto, M. (2004) Biphenyl dioxygenases: functional versatilities and directed evolution. *J. Bacteriol. 186*, 5189-5196.

(6) Demaneche, S., Meyer, C., Micoud, J., Louwagie, M., Willison, J. C., and Jouanneau, Y. (2004) Identification and functional analysis of two aromatic ring-hydroxylating dioxygenases from a *Sphingomonas* strain degrading various polycyclic aromatic hydrocarbons. *Appl. Environ. Microbiol. 70*, 6714-6725.

(7) Krivobok, S., Kuony, S., Meyer, C., Louwagie, M., Willison, J. C., and Jouanneau, Y. (2003) Identification of pyrene-induced proteins in *Mycobacterium* sp. 6PY1 : Evidence for two ring-hydroxylating dioxygenases. *J. Bacteriol. 185*, 3828-3841.

(8) Saito, A., Iwabuchi, T., and Harayama, S. (2000) A novel phenanthrene dioxygenase from *Nocardioides* sp strain KP7: Expression in *Escherichia coli*. *J. Bacteriol. 182*, 2134-2141.

(9) Furukawa, K. (2000) Engineering dioxygenases for efficient degradation of environmental pollutants. *Curr. Opin. Biotechnol. 11*, 244-249.

(10) Wolfe, M. D., Parales, J. V., Gibson, D. T., and Lipscomb, J. D. (2001) Single turnover chemistry and regulation of O-2 activation by the oxygenase component of naphthalene 1,2-dioxygenase. *J. Biol. Chem. 276*, 1945-1953.

(11) Batie, C. J., LaHaie, E., and Ballou, D. P. (1987) Purification and characterization of phthalate oxygenase and phthalate oxygenase reductase from *Pseudomonas cepacia*. *J. Biol. Chem. 262*, 1510-1518.

(12) Correll, C. C., Batie, C. J., Ballou, D. P., and Ludwig, M. L. (1992) Phthalate dioxygenase reductase: a modular structure for electron transfer from pyridine nucleotides to [2Fe-2S]. *Science 258*, 1604-1610.

(13) Kauppi, B., Lee, K., Carredano, E., Parales, R. E., Gibson, D. T., Eklund, H., and Ramaswamy, S. (1998) Structure of an aromatic-ring-hydroxylating dioxygenase-naphthalene 1,2- dioxygenase. *Structure 6*, 571-586.

(14) Parales, R. E., Lee, K., Resnick, S. M., Jiang, H. Y., Lessner, D. J., and Gibson, D. T. (2000) Substrate specificity of naphthalene dioxygenase: Effect of specific amino acids at the active site of the enzyme. *J. Bacteriol. 182*, 1641-1649.

(15) Furusawa, Y., Nagarajan, V., Tanokura, M., Masai, E., Fukuda, M., and Senda, T. (2004) Crystal structure of the terminal oxygenase component of biphenyl dioxygenase derived from *Rhodococcus* sp. strain RHA1. *J. Mol. Biol. 342*, 1041-1052.

(16) Kanaly, R. A., and Harayama, S. (2000) Biodegradation of high-molecular-weight polycyclic aromatic hydrocarbons by bacteria. *J. Bacteriol. 182*, 2059-2067.

(17) IARC. (1983) Polynuclear Aromatic Compounds, Part 1: Chemical, Environmental and Experimental Data, in *Monographs on the evaluation of carcinogenic risks to humans*.

(18) Willison, J. C. (2004) Isolation and characterization of a novel sphingomonad capable of growth with chrysene as sole carbon and energy source. *FEMS Microbiol. Lett. 241*, 143-150.

(19) Sambrook, J., Fritsch, E. F., and Maniatis, T. (1989) *Molecular cloning : a laboratory manual*, 2nd ed., Cold Spring Harbor Laboratory Press, Cold Spring Harbor, N. Y.

(20) Hurtubise, Y., Barriault, D., Powlowski, J., and Sylvestre, M. (1995) Purification and characterization of the *Comamonas testosteroni* B-356 biphenyl dioxygenase components. *J. Bacteriol. 177*, 6610-6618.

(21) Rodarie, D., and Jouanneau, Y. (2001) Genetic and biochemical characterization of the biphenyl dioxygenase from *Pseudomonas* sp. strain B4. *J. Microbiol. Biotechnol. 11*, 762-771.

(22) Jouanneau, Y., and Meyer, C. (2006) Purification and characterization of an arene *cis*-dihydrodiol dehydrogenase endowed with broad substrate specificity toward polycyclic aromatic hydrocarbon dihydrodiols. *Appl. Environ. Microbiol. 72*, 4726-4734.

(23) Blair, D., and Diehl, H. (1961) Bathophenantroline disulphonic acid and bathocuproine disulphonic acid, water soluble reagents for iron and copper. *Talanta 7*, 163-174.

(24) Bradford, M. M. (1976) A rapid and sensitive method for the quantitation of microgram quantities of protein utilizing the principle of protein-dye binding. *Anal. Biochem. 72*, 248-254.

(25) Pelley, J. W., Garner, C. W., and Little, G. H. (1978) A simple rapid buiret method for the estimation of protein in samples containing thiols. *Anal. Biochem. 86*, 341-343.

(26) Jouanneau, Y., Meyer, C., Naud, I., and Klipp, W. (1995) Characterization of an *fdxN* mutant of *Rhodobacter capsulatus* indicates that ferredoxin I serves as electron donor to nitrogenase. *Biochim. Biophys. Acta 1232*, 33-42.

(27) Naud, I., Vincon, M., Garin, J., Gaillard, J., Forest, E., and Jouanneau, Y. (1994) Purification of a sixth ferredoxin from *Rhodobacter capsulatus*. Primary structure and biochemical properties. *Eur. J. Biochem. 222*, 933-939.

(28) Jeffrey, A. M., Yeh, H. J., Jerina, D. M., Patel, T. R., Davey, J. F., and Gibson, D. T. (1975) Initial reactions in the oxidation of naphthalene by *Pseudomonas putida*. *Biochemistry 14*, 575-584.

(29) Jerina, D. M., Selander, H., Yagi, H., Wells, M. C., Davey, J. F., Mahadevan, V., and Gibson, D. T. (1976) Dihydrodiols from anthracene and phenanthrene. *J. Am. Chem. Soc. 98*, 5988-5996.

(30) Boyd, D. R., Sharma, N. D., Agarwal, R., Resnick, S. M., Schocken, M. J., Gibson, D. T., Sayer, J. M., Yagi, H., and Jerina, D. M. (1997) Bacterial dioxygenase-catalysed dihydroxylation and chemical resolution routes to enantiopure *cis*-dihydrodiols of chrysene. *J. Chem. Soc. Perkin Trans. 1*, 1715-1723.

(31) Gibson, D. T., Mahadevan, V., Jerina, D. M., Yogi, H., and Yeh, H. J. (1975) Oxidation of the carcinogens benzo [a] pyrene and benzo [a] anthracene to dihydrodiols by a bacterium. *Science 189*, 295-297.

(32) Jerina, D. M., Vanbladeren, P. J., Yagi, H., Gibson, D. T., Mahadevan, V., Neese, A. S., Koreeda, M., Sharma, N. D., and Boyd, D. R. (1984) Synthesis and absolute-configuration of the bacterial *cis*-1,2-dihydrodiol, *cis*-8,9-dihydrodiol, and *cis*-10,11-dihydrodiol metabolites of benz[a]anthracene formed by a strain of *Beijerinckia*. *J. Org. Chem. 49*, 3621-3628.

(33) Lee, K. (1999) Benzene-induced uncoupling of naphthalene dioxygenase activity and enzyme inactivation by production of hydrogen peroxide. *J. Bacteriol. 181*, 2719-2725.

(34) Selifonov, S. A., Grifoll, M., Eaton, R. W., and Chapman, P. J. (1996) Oxidation of naphthenoaromatic and methyl-substituted aromatic compounds by naphthalene 1,2-dioxygenase. *Appl. Environ. Microbiol. 62*, 507-514.

(35) Rehmann, K., Hertkorn, N., and Kettrup, A. A. (2001) Fluoranthene metabolism in *Mycobacterium* sp strain KR20: identity of pathway intermediates during degradation and growth. *Microbiology-Sgm 147*, 2783-2794.

(36) Gibson, D. T. (1999) *Beijerinckia* sp strain B1: a strain by any other name. *J.Ind. Microbiol. Biotechnol. 23*, 284-293.

(37) Batie, C. J., Ballou, D. P., and Correll, C. C. (1992) Phthalate dioxygenase reductase and related flavin-iron-sulfur containing electron transferases, in *Chemistry and Biochemistry of Flavoenzymes* (Müller, F., Ed.) pp 543-556, CRC Press, Boca Raton, Fla.

(38) Eby, D. M., Beharry, Z. M., Coulter, E. D., Kurtz, D. M., Jr., and Neidle, E. L. (2001) Characterization and evolution of anthranilate 1,2-dioxygenase from *Acinetobacter* sp. strain ADP1. *J. Bacteriol. 183*, 109-118.

(39) Imbeault, N. Y., Powlowski, J. B., Colbert, C. L., Bolin, J. T., and Eltis, L. D. (2000) Steady-state kinetic characterization and crystallization of a polychlorinated biphenyl-transforming dioxygenase. *J. Biol. Chem. 275*, 12430-12437.

(40) Ensley, B. D., and Gibson, D. T. (1983) Naphthalene dioxygenase: purification and properties of a terminal oxygenase component. *J. Bacteriol. 155*, 505-11.

(41) Hurtubise, Y., Barriault, D., and Sylvestre, M. (1996) Characterization of active recombinant his-tagged oxygenase component of *Comamonas testosteroni* B-356 biphenyl dioxygenase. *J. Biol. Chem. 271*, 8152-8156.

(42) Parales, R. E., Huang, R., Yu, C. L., Parales, J. V., Lee, F. K., Lessner, D. J., Ivkovic-Jensen, M. M., Liu, W., Friemann, R., Ramaswamy, S., and Gibson, D. T. (2005) Purification, characterization, and crystallization of the components of the nitrobenzene and 2-nitrotoluene dioxygenase enzyme systems. *Appl. Environ. Microbiol. 71*, 3806-3814.

(43) Wang, R. F., Wennerstrom, D., Cao, W. W., Khan, A. A., and Cerniglia, C. E. (2000) Cloning, expression, and characterization of the *katG* gene, encoding catalase-peroxidase, from the polycyclic aromatic hydrocarbon-degrading bacterium *Mycobacterium* sp. strain PYR-1. *Appl. Environ. Microbiol. 66*, 4300-4304.

(44) Mahaffey, W. R., Gibson, D. T., and Cerniglia, C. E. (1988) Bacterial oxidation of chemical carcinogens: formation of polycyclic aromatic acids from benz[a]anthracene. *Appl. Environ. Microbiol. 54*, 2415-2423.

(45) Boyd, D. R., Sharma, N. D., Hempenstall, F., Kennedy, M. A., Malone, J. F., Allen, C. C. R., Resnick, S. M., and Gibson, D. T. (1999) *bis*-*cis*-Dyhydrodiols: a new class of metabolites resulting from biphenyl dioxygenase-catalyzed sequential asymetric *cis*-dihydroxylation of polycyclic arenes and heteroarenes. *J. Org. Chem. 64*, 4005-4011.

Table 1 : Specific activity and iron content of different preparations of ht-PhnI .
Iron was determined by chemical analysis and EPR spectroscopy. The standard error for each determination was less than 10%.

| Ht-PhnI preparation | 1 | 2 | 3 |
|---|---|---|---|
| Concentration[a] (µM) | 21.0 | 25.7 | 19.7 |
| Total Fe (atoms/αβ) | 1.73 | 2.54 | 2.55 |
| Fe(II)-NO (spin/αβ) | 0.20 | 0.78 | 0.92 |
| [2Fe-2S] (spin/αβ) | 0.75 | 0.85 | 0.75 |
| Specific activity [b] (U/mg) | 0.41 ± 0.035 | 0.86 ± 0.01 | 0.58 ± 0.01 |

[a] The protein concentration were estimated from the microbiuret assays, using a molecular mass of 215 kDa for ht-PhnI. Based on these determinations, the absorbance coefficient of ht-PhnI at 458 nm was calculated to be $\varepsilon_{458}$ = 12,500 $M^{-1}.cm^{-1}$

[b] as determined by the NADH oxidation assay with naphthalene as substrate. The molar ratio PnA3/PhnI was approx. 9.5 in these assays.

Table 2 : Specific activity and coupling efficiencies of the dioxygenase as a function of PAH substrates[a]

| Substrate | $O_2$ consumption | Dihydrodiol formed | Dihydrodiol/$O_2$ [b] | $H_2O_2$/ $O_2$ | NADH oxidation | Diol/NADH |
|---|---|---|---|---|---|---|
| | nmol.min$^{-1}$.mg$^{-1}$ | nmol.min$^{-1}$.mg$^{-1}$ | | | nmol.min$^{-1}$.mg$^{-1}$ | |
| Naphthalene | 460 (20) | 418 (20) | 1.03 (0.07) | 0.122 (0.010) | 510 (50) | 0.92 (0.06) |
| Biphenyl | 410 (40) | 260 (12) | 0.70 (0.01) | 0.28 (0.015) | 290 (35) | 0.67 (0.03) |
| Phenanthrene | 370 (30) | 300 (30) | 0.83 (0.03) | 0.128 (0.028) | 465 (10) | 0.46 (0.01) |
| Anthracene | 360 (30) | 130 (10) | 0.36 (0.02) | 0.53 (0.02) | 141 (6) | 0.62 (0.01) |
| Fluorene | 91.5 (7) | 91 (5) [c] | 0.99 (0.05) | - | - | - |
| Fluoranthene | 147 (24) | 76 (5) [c] | 0.51 (0.04) | - | - | - |
| Benz[a]anthracene | 167 (20) | 52 (5) [d] | 0.31 (0.04) | 0.48 (0.04) | 46 (2) | 0.40 (0.07) |
| Chrysene | 49 (2) | 2.5 (0.3) | 0.082 (0.011) | 0.59 (0.08) | 9.6 (0.6) | 0.098 (0.007) |
| Benzo[a]pyrene | 43 (8) | 7.8 (0.7) | 0.33 (0.02) | 0.50 (0. 005) | 10 (1) | 0.19 (0.01) |
| 1,2-Benz[a]anthracene diol | 2500 (400) | 11 (2) | 0.0044 (0.001) | 0.83 | - | - |
| 10,11-Benz[a]anthracene diol | 1500 (150) | 16.5 (0.5) | 0.011 (0.0005) | 0.66 | - | - |

[a] The indicated values represent means obtained from two to four determinations, with standard deviations given in parentheses. - means not determined. $O_2$ consumption and NADH oxidation represent initial rates corrected for the background activity observed in the absence of PAH

substrate. The rates of dihydrodiol formation are average rates calculated over the duration of the assay which lasted between 4 min(naphthalene) and 11 min (benzo[a]pyrene).

[b] Values were calculated as ratios between the rates of dihydrodiol formation (column 3), and the average rates of $O_2$ consumption over the duration of the assay, which are lower than the initial rates of $O_2$ consumption given in column 2.

[c] These values represent rates of substrate oxidation, because the oxidation products from florene and fluoranthene could not be measured accurately (see text).

[d] Three diol isomers were produced, which were not separated by HPLC and quantified as a mixture (see Methods). The contribution of *bis*-*cis*-dihydrodiol, also produced in the reaction, was not taken into account in the calculations.

Table 3 : GC-MS identification of PAH oxidation products formed by CHY-1 dioxygenase.

| Substrate | Properties of trimethylsilyl derivates of products | | | Identification |
|---|---|---|---|---|
| | GC retention time (min) | Relative peak area (%) | m/z and relative abundance (%) of major fragments | |
| Fluorene | 15.18 | 6 | 254($M^+$, 41), 239(8), 166(15), 165(100) | 9-Fluorenol[a] |
| | 16.18 | 6 | 254($M^+$, 100), 239(39), 224(19), 223(89), 178(8), 165(13) | Hydroxyfluorene |
| | 16.49 | 14 | 254($M^+$, 100), 239(70), 223(3), 195(2.5), 178(11), 165(63) | Hydroxyfluorene |
| | 16.67 | 16 | 254($M^+$, 100), 239(67), 223(2), 195(2), 178(10), 165(25) | Hydroxyfluorene |
| | 17.70 | 58 | 342($M^+$, 100), 327(4), 253(62), 238(5), 178(5), 164(5), 163(5) | Dihydroxyfluorene |
| Fluoranthene | 20.74 | 100 | 290($M^+$, 100), 275(70), 219(11), 215(11), 201(14), 200(14), 189(15) | 8-Hydroxyfluoranthene |
| Chrysene | 23.08 | >90[b] | 406($M^+$, 28), 317(14), 316(18), 303(32), | 3,4-dihydrodiol[c] |
| | | | 244(14), 228(40), 226(33), 215(66), 191(100) | |
| | 24.55 | <10[b] | 584($M^+$, 8), 393(8), 355(8), 282(13), 281(40), 207(100), 191(73) | 3,4,9,10-*bis*-dihydrodiol[c] |
| Benz[a]anthracene | 22.65 | 68 | 406($M^+$, 28), 316(26), 303(30), 281(14), 228(28), 226(25), 215(36), 191(50), 73(100) | 1,2-Dihydrodiol[c] |
| | 22.95 | 23 | 406($M^+$, 32), 316(29), 303(59), 228(21), 226(14), 191(46), 73(100) | 10,11-Dihydrodiol[c] |
| | 23.43 | 9 | 406($M^+$, 31), 316(27), 303(80), 281(16), 228(18), 226(14), 215(29), 191(68), 73(100) | 8,9-Dihydrodiol[c] |
| | 23.53 | 20-40 [d] | 584($M^+$, 46), 481(21) 392(9), 355(16), 281(22), 207(48), 191(100) | 1,2,10,11-*bis*-Dihydrodiol[c] |
| Benzo[a]pyrene | 25.88 | 100 | 430($M^+$, 5.5), 415 (1.5), 341(5), 327(8), 281(20), 252(8), 207(100) | 9,10-Dihydrodiol[c] |

[a] Identification based on match of mass spectrum and GC retention time with those of an authentic sample.

[b] Percentages estimated from HPLC peak area.

[c] Identification based on comparisons of the GC-MS and UV absorbance data of the products with previously published data (see text).

[d] The indicated range represents an estimation of the *bis*-dihydrodiol/dihydrodiols ratio, as indicated in the text.

Table 4 : Solvent-facilitated formation of benz[a]anthracene *cis*-dihydrodiols catalyzed by PhnI in single turnover reactions.

| Acetonitrile % in the reaction | Benz[a]anthracene solubilized (µM) | Incubation time (min) | Total diols/PhnI | Percentage of dihydrodiols produced as[a] 1,2-diol | 10,11-diol | 8,9-diol |
|---|---|---|---|---|---|---|
| 0 | 0.025 | 1 | 0.44 | | | |
| | | 10 | 2.12[b] | 28 | 64 | 8 |
| 20 | 2.5 | 1 | 2.01 | 30 | 53 | 17 |
| | | 10 | 1.94 | 36 | 50 | 14 |
| 30 | 27.0 | 1 | 1.12 | 29 | 43 | 28 |
| | | 10 | 0.96 | 28.5 | 54 | 17.5 |

[a] As determined from the peak area of trimethylsilyl derivates analyzed by GC-MS

[b] Some *bis*-*cis*-dihydrodiol was also produced which accounted for about 1.5% of the total diols formed, based on HPLC determination.

**Figure legends**

FIGURE 1: SDS-PAGE of the purified components of the ring-hydroxylating dioxygenase of *Sphingomonas* strain CHY-1. Polypeptides were electrophoresed on a 15% polyacrylamide slab gel. Lane 1: molecular mass markers. Lane 2: ht-PhnI, 3.2 µg. Lane 3: ht-PhnA3, 1.0 µg. Lane 4: ht-PhnA4, 1.8 µg.

FIGURE 2 : Dioxygenation reactions of four-ring PAHs catalyzed by ht-PhnI.

Benz[a]anthracene was converted into three dihydrodiols isomers (a), in proportions which varied depending on experimental conditions (see text). The 1,2- and 10,11-isomers were subjected to a second dihydroxylation, yielding the same *bis-cis*-dihydrodiol (b). Chrysene was oxidized to a single dihydrodiol, which could subsequently react with the dioxygenase to yield the 3,4,9,10-*bis-cis*-dihydrodiol (c). Reaction products were identified as indicated in the text. Stereochemical configurations were assumed to be identical to those reported in previous studies on the *S. yanoikuyae* enzyme (*32, 45*).

FIGURE 3 : EPR spectra of nitrosyl complexes of ht-PhnI in the presence or absence substrates.

Protein samples contained 22.0 µM ht-PhnI in 0.18 ml of 50 mM potassium phosphate, pH 7.5, and, either no substrate (spectrum a) or one of the following substrates (0.1 mM) added in 10 µl acetonitrile : 1,2-benz[a]anthracene dihydrodiol (spectrum b); 10,11-benz[a]anthracene dihydrodiols (spectrum c); benz[a]anthracene (spectrum d); naphthalene (spectrum d). After 15 min at room temperature under argon, nitrosyl complexes were prepared (see Materials and Methods), samples were transferred into EPR tubes and frozen. Acquisition conditions : Temperature; 4K; Microwave power: 250 µW; modulation frequency: 100 kHz; modulation amplitude: 1mT. Relevant *g* values are indicated.

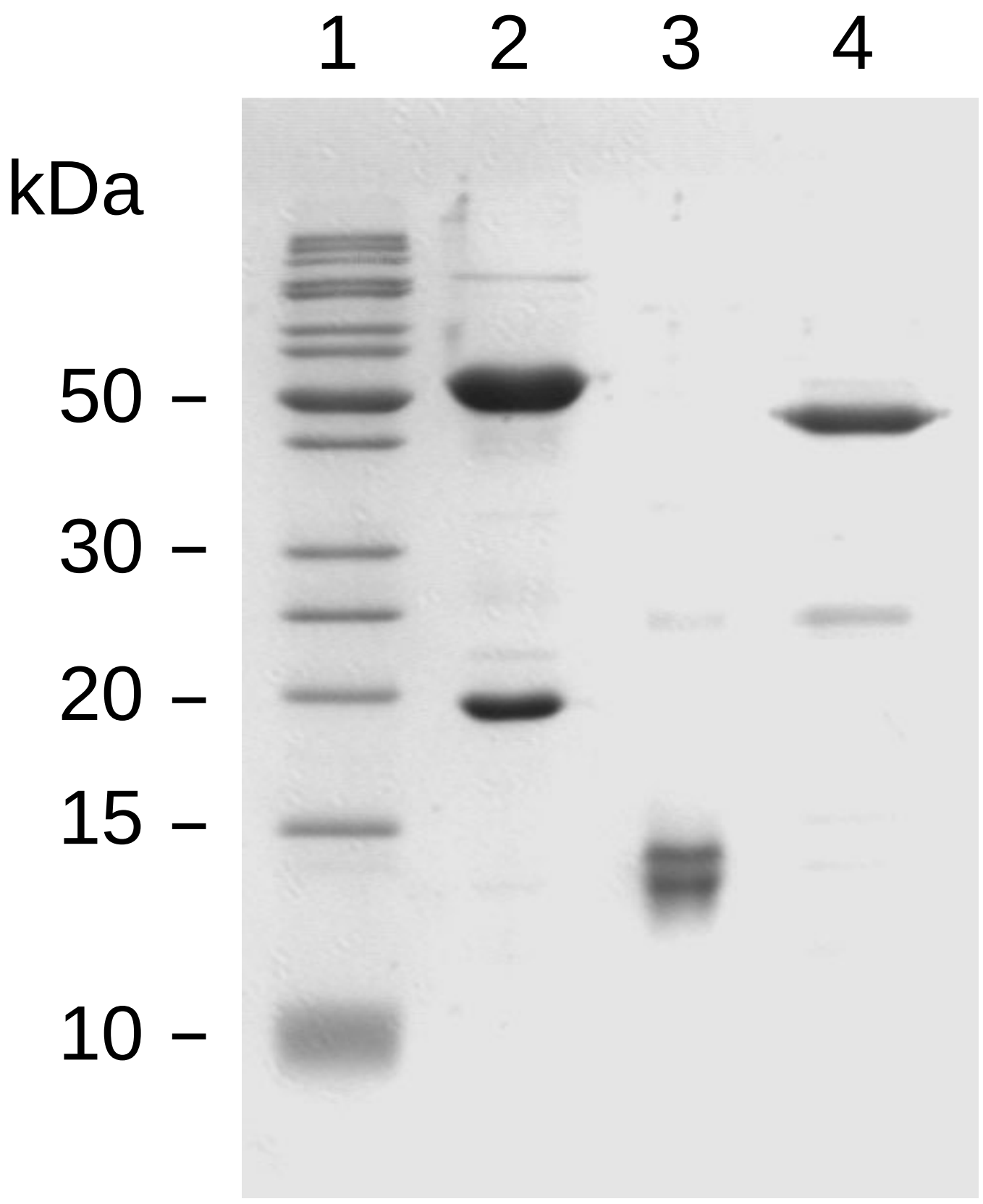

Fig. 1

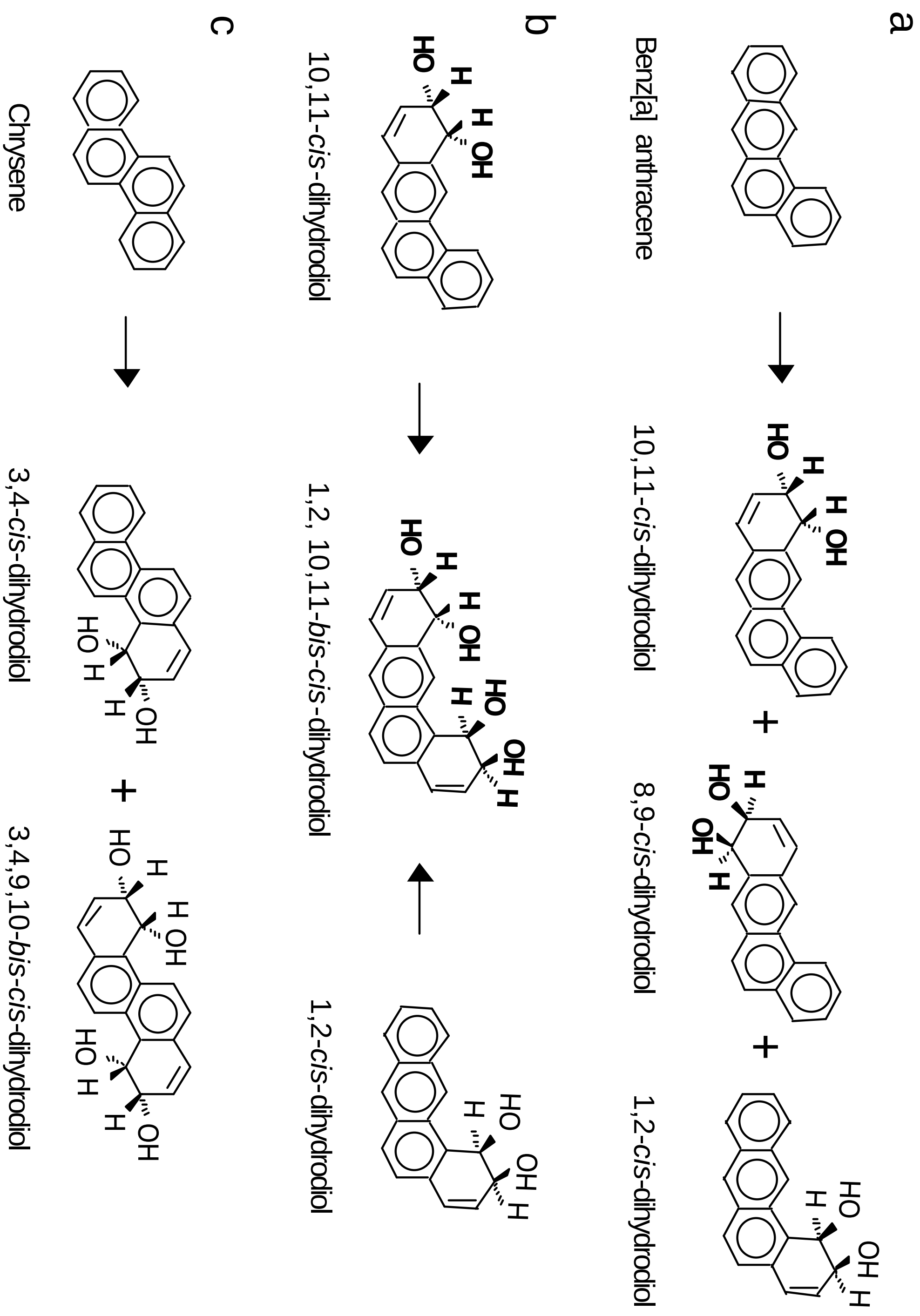

Fig. 2

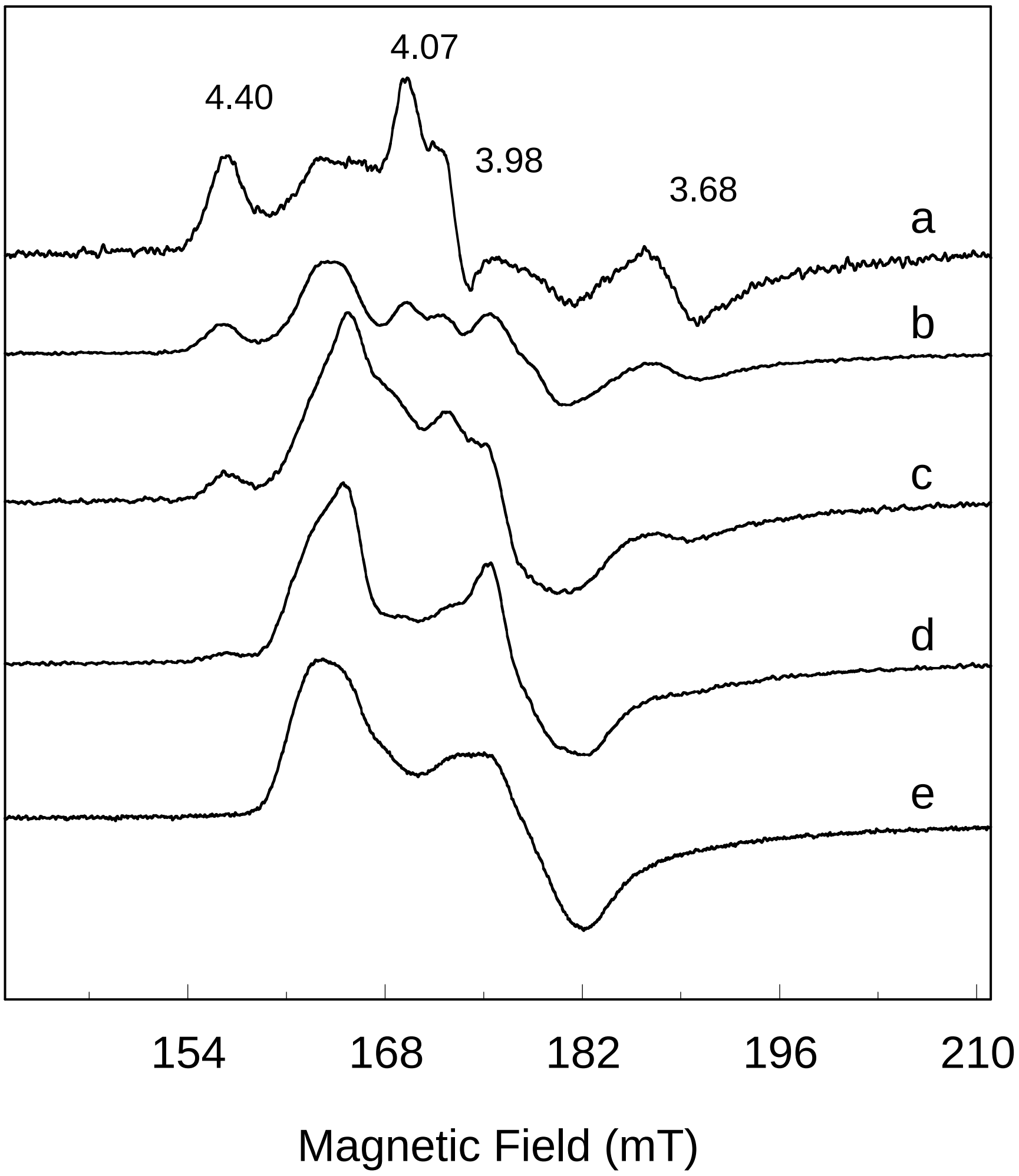

Fig. 3

For Table of Contents Use Only

**Characterization of a naphthalene dioxygenase endowed with an exceptionally broad specificity towards polycyclic aromatic hydrocarbons†**

By Yves Jouanneau *, ‡, Christine Meyer‡, Jean Jakoncic§, Vivian Stojanoff§, Jacques Gaillard||

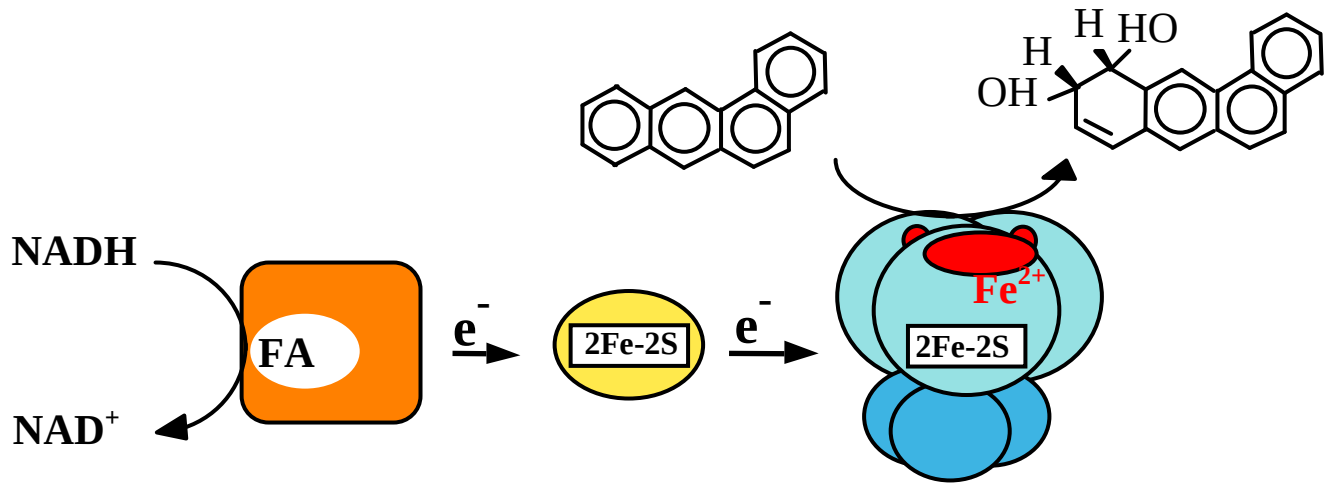